\documentclass[%
 reprint,
superscriptaddress,
 amsmath,amssymb,
 aps,
]{revtex4-2}

\usepackage{graphicx}
\usepackage{dcolumn}
\usepackage{bm}
\usepackage{hyperref}
\usepackage[mathlines]{lineno}
\usepackage{amsmath}
\usepackage{graphicx}
\usepackage{subfigure}
\usepackage{float}

\begin{document}

\preprint{APS/123-QED}

\title{A Kalman-smoother based data imputation strategy to data gaps in spaceborne gravitational wave detectors}

\author{Tingyang Shen}
\email{shentinagyang16@mail.ucas.ac.cn}
\affiliation{University of Chinese Academy of Sciences (UCAS), Beijing 100049, China}
\affiliation{International Centre for Theoretical Physics Asia-Pacific (ICTP-AP), UCAS, Beijing 100190, China}
\affiliation{Taiji Laboratory for Gravitational Wave Universe (Beijing/Hangzhou), UCAS, Beijing 100190, China}

\author{He Wang}
\email{hewang@ucas.ac.cn}
\affiliation{University of Chinese Academy of Sciences (UCAS), Beijing 100049, China}
\affiliation{International Centre for Theoretical Physics Asia-Pacific (ICTP-AP), UCAS, Beijing 100190, China}
\affiliation{Taiji Laboratory for Gravitational Wave Universe (Beijing/Hangzhou), UCAS, Beijing 100190, China}

\author{Jibo He}
\email{jibo.he@ucas.ac.cn}
\affiliation{University of Chinese Academy of Sciences (UCAS), Beijing 100049, China}
\affiliation{International Centre for Theoretical Physics Asia-Pacific (ICTP-AP), UCAS, Beijing 100190, China}
\affiliation{Taiji Laboratory for Gravitational Wave Universe (Beijing/Hangzhou), UCAS, Beijing 100190, China}
\affiliation{Hangzhou Institute for Advanced Study, UCAS, Hangzhou 310024, China}

\date{\today}

\begin{abstract}
Massive black hole binaries (MBHBs) and other sources within the frequency band of spaceborne gravitational wave observatories like the Laser Interferometer Space Antenna (LISA), Taiji and Tianqin pose unique challenges, as gaps and glitches during the years-long observation lead to both loss of information and spectral leakage. We propose a novel data imputation strategy based on Kalman filter and smoother to mitigate gap-induced biases in parameter estimation. Applied to a scenario where traditional windowing and smoothing technique introduce significant biases, our method mitigates the biases and demonstrates lower computational cost compared to existing data augmentation techniques such as noise inpainting. This framework presents a new gap treatment approach that balances robustness and efficiency for space-based gravitational wave data analysis.
\end{abstract}

\maketitle


\section{Introduction}
Since the landmark detection of GW150914~\cite{PhysRevLett.116.061102,PhysRevLett.116.241103}, global interest in gravitational wave (GW) astronomy has been growing, leading to the initiation of various projects aimed at detecting gravitational wave signals across different frequency bands. Ground based detectors like the Laser Interferometer Gravitational Wave Observatory (LIGO)~\cite{LIGO2015}, Virgo~\cite{Acernese_2014} and the Kamioka Gravitational Wave Detector (KAGRA)~\cite{Kagra2019} focus mainly on a sensitive range between $\sim$10~Hz to a few kHz~\cite{PhysRevX.9.031040}. In order to detect sources from lower frequency bands such as Massive Black hole binaries (MBHB)~\cite{Sesana_2005}, extreme mass ratio inspirals (EMRI)~\cite{Gair_2017}, stellar-mass black hole inspirals~\cite{PhysRevD.102.124037} and stochastic gravitational wave background (SGWB)~\cite{Bartolo_2018}, multiple spaceborne gravitational wave missions have been proposed with different configurations and characteristics. LISA and Taiji missions share a similar configuration as a three-satellite constellation trailing~\cite{colpi2024lisadefinitionstudyreport} and leading~\cite{10.1093/nsr/nwx116} the earth respectively in the heliocentric orbit, each with their own armlengths of 2.5 million km for LISA~\cite{colpi2024lisadefinitionstudyreport} and 3 million km for Taiji~\cite{10.1093/nsr/nwx116}, resulting in slight differences on the sensitivity. On the other hand, the geocentric mission TianQin has an armlength of 0.17 million km, and to mitigate the thermal load, TianQin adopts a 3-months-on and 3-months-off observation scheme~\cite{Luo_2016}, giving rise to unique challenges compared to the heliocentric missions like LISA and Taiji.

Compared to existing GW sources for ground-based detectors like LIGO and Virgo, sources for spaceborne GW missions, primarily MBHBs, would remain in observable bandwidth for longer time, from hours to years. As a result, data gaps emerge as a primary data analysis challenge for future missions. Gaps in observations can arise from several factors which are generally categorized into two categories: scheduled and unscheduled gaps. 

Scheduled gaps are interruptions in observations due to pre-planned maintenance circles or mandatory downtime of observation systems for various reasons, many identified factors includes antenna re-pointing (gap lasts 3 hours every 14 days), tilt-to-length coupling constant estimation (gap lasts 2 days and has a frequency of about four times per year) and point-ahead angle mechanism (PAAM) adjustments (three times per day, lasting about 100 seconds each)~\cite{burke2025mindgapaddressingdata}. In general, scheduled gaps are periods during which accumulative effects are addressed and the whole system go through maintenance and potential reparation (especially in ground-based missions). 

Unscheduled gaps, on the other hand, are mainly caused by various unforeseeable events that will eventually translate into scenarios where data quality is severely compromised: For spaceborne missions like LISA, Taiji and TianQin, unforeseeable problems in instruments and random physical events like micrometeorite collisions can render the detectors inoperable. Such collisions are estimated to occur approximately 30 times per year and each event can cause a data gap lasting up to one day~\cite{burke2025mindgapaddressingdata}. Other minor problems like instrument outages can also contribute to unscheduled gaps. Also, nonstationary events known as glitches have been observed in various GW missions, and sereval schemes have been devised to characterize and mitigate such effects. Windowing such events out and treating them as gaps also emerges as a fallback solution to glitches, as is shown in Ref~\cite{Castelli_2025}. These potential applications further highlight the importance of developing a comprehensive strategy for gap treatment in general.

Since the launch of LISA Pathfinder mission, many efforts have been made to evaluate the effects of different patterns of gaps on data analysis. Carré and Porter found that untreated gaps on time domain would lead to significant spectral leakage on Fourier domain hindering data analysis~\cite{carré2010effectdatagapslisa}. It is suggested in Ref.~\cite{Dey_2021} that when treated with smoothing technique, unscheduled gaps tend to have greater impact on the detectability of signals than scheduled gaps and such gaps will induce significant loss of SNR (signal-to-noise ratio), and the closer gaps are to the merger, the greater their influence would be. The gap happens during the merger phase can severely impact parameter estimation performance and in many cases result in loss of detection~\cite{Dey_2021} completely. 

To counter such adverse effects, many techniques have been developed. These techniques are generally divided into two categories: data augmentation and windowing. Broadly speaking, data augmentation methods~\cite{Baghi_2019,Blelly_2021,Wang_2025} use Bayesian techniques to estimate and sample parameters for both the signal and noise background, thereby reconstructing the missing data. However, this approach considerably increases computational cost and is typically applied to Galactic Binary sources rather than MBHB systems. Applying the same approach to MBHB signals would further escalate computational demands due to the cost of template generation and parameter estimation. More recent works in this category, such as noise inpainting~\cite{Wang_2025} and autoencoder-based methods~\cite{Mao_2025}, have been applied to MBHB cases and attempt to mitigate computational costs in different ways.

Windowing, on the other hand, addresses spectral leakage by applying a window function to both sides of the gap~\cite{Baghi_2019}. However, this treatment leads to the loss of additional valid data and renders the resulting data non-stationary, making noise mismodeling a significant issue. Recent work~\cite{burke2025mindgapaddressingdata} analyzes this problem and proposes a formalism to characterize the resulting modeling errors.

In this work, we propose a new data imputation technique based on the Kalman smoother in order to repair gaps and improve the posterior estimation performance. The Kalman smoother is a widely-used technique across many fields and can handle noisy observations without the need to provide labeled training data. Its computational cost remains relatively low compared to previous data augmentation techniques, despite having a computational complexity that scales cubically with the size of the state space. It also makes use of existing data points which includes both data and signal components rather than noise parameters only to generate inpainting data points, which may retain more relevant information. In our testing scenario, it is found that the Kalman smoother can successfully handle randomized gap patterns as long as such gaps are not dangerously close to time of coalescence.

This paper is organized as follows: Sec.II introduces our methodologies, including a brief introduction on data generation in Sec.II A, gap introduction and treatment in Sec.II B as well as principles for Bayesian estimation and sampling techniques in Sec.II C. Sec.III shows our results with details about our injection and noise realization in Sec.III A and parameter estimation results for our injected case in Sec.III B with comparison between standard windowing and Kalman-based treatment. We will then summarize our work in Sec.IV. 

\section{Methodology}
\subsection{Data preparation}
In data generation process we made use of existing GPU-accelerated Black Hole Binary Waveforms package bbhx~\cite{Katz_2020, Katz_2022} to generate our waveform for testing. We use the following set of parameters to describe our signal: chirp mass $\mathcal{M}_c$, mass ratio $q$, time of coalescence of the binary $t_c$, luminosity distance $D_L$, spin $\chi_1$ and $\chi_2$, inclination $\iota$, ecliptic longitude $\lambda$, ecliptic latitude $\beta$, polarization angle $\psi$ and coalescence phase $\phi_c$.
We generated our waveform in IMRPhenomD~\cite{PhysRevD.93.044006} and put it through the Time Delay Interferometery (TDI) channel of A, E and T to obtain the noise-free signal in that three channels.

For the noise term $n_{ij}$ where $ij$$\in\{12,23,31,21,32,13\} $, $i$ and $j$ specifying the index of S/C receiving and emitting lasers, respectively. The noise here is seen as a combination of various noise factors including read-out noise, test-mass acceleration noise, optical path noise, etc. The dominant contribution is described as a combination of the optical metrology system (OMS) noises $N_{ij}$ and test-mass acceleration (ACC) noises $\delta_{ij}$~\cite{Quang_Nam_2023},
\begin{equation}
\begin{aligned}
n_{i j}(t)=N_{i j}(t)+\delta_{i j}(t)+\mathbf{D}_{i j} \delta_{j i}(t).
\end{aligned}
\end{equation}
The delay operator $\mathbf{D}_{i j}$ is defined as $\mathbf{D}_{i j} f(t) \equiv$ $f\left(t-d_{i j}(t)\right)$, when acted on an arbitrary function of time $f(t)$. According to the baseline design of Taiji, the nominal Power Spectral Density (PSD) of $N_{i j}$ and $\delta_{i j}$ take the forms of
\begin{equation}
\begin{aligned}
S_{i j, \mathrm{OMS}}(f)= & A_{i j, \mathrm{OMS}}^2\left(\frac{2 \pi f}{c}\right)^2\left[1+\left(\frac{2 \mathrm{mHz}}{f}\right)^4\right], \\
S_{i j, \mathrm{ACC}}(f)= & A_{i j, \mathrm{ACC}}^2\left(\frac{1}{2 \pi f c}\right)^2\left[1+\left(\frac{0.4 \mathrm{mHz}}{f}\right)^2\right] \\
& \times\left[1+\left(\frac{f}{8 \mathrm{mHz}}\right)^4\right],
\end{aligned}
\end{equation}
where $A_{i j, \mathrm{OMS}} \equiv 8 \times 10^{-12} \mathrm{~m} / \sqrt{\mathrm{Hz}}$ and $A_{i j, \mathrm{ACC}} \equiv 3 \times$ $10^{-15} \mathrm{~m} / \mathrm{s}^2 / \sqrt{\mathrm{Hz}}$. 

In order to suppress laser frequency noise, we used TDI observables and these observables are defined as,
\begin{equation}
\mathrm{TDI}=\sum_{i j \in \mathcal{I}_2} \mathbf{P}_{i j} \eta_{i j}.
\end{equation}
For the second generation TDI observable $X_2$, $P_{ij}$ is defined as~\cite{Wang_2022},
\begin{equation}
\begin{aligned}
& \mathbf{P}_{12}=1-\mathbf{D}_{131}-\mathbf{D}_{13121}+\mathbf{D}_{1213131}, \\
& \mathbf{P}_{23}=0, \\
& \mathbf{P}_{31}=-\mathbf{D}_{13}+\mathbf{D}_{1213}+\mathbf{D}_{121313}-\mathbf{D}_{13121213}, \\
& \mathbf{P}_{21}=\mathbf{D}_{12}-\mathbf{D}_{1312}-\mathbf{D}_{131212}+\mathbf{D}_{12131312}, \\
& \mathbf{P}_{32}=0, \\
& \mathbf{P}_{13}=-1+\mathbf{D}_{121}+\mathbf{D}_{12131}-\mathbf{D}_{1312121}.
\end{aligned}
\end{equation}
Definitions for observable $Y_2$ and $Z_2$ can also be obtained according to their TDI scheme following the permutation rule of 1 to 2, 2 to 3 and 3 to 1. In order to obtain the signal channel A and E and the null channel T, we follow this definition~\cite{PhysRevD.66.122002},
\begin{equation}
\begin{aligned}
A_2 & =\frac{Z_2-X_2}{\sqrt{2}}, \\
E_2 & =\frac{X_2-2 Y_2+Z_2}{\sqrt{6}}, \\
T_2 & =\frac{X_2+Y_2+Z_2}{\sqrt{3}}.
\end{aligned}
\end{equation}
Then we combine our noise-free signal with the noise drawn from the above-mentioned PSD on A, E and T channel to obtain our full observation data. In practice, we make use of the PSD and noise generation functions also used in data generation for Taiji Data Challenge~\cite{du2025realisticdetectionpipelinestaiji}.

\subsection{Gap introduction and treatment}
Gaps are generated using gap indices from randomized or arbitrary input, and the corresponding time-domain data points are set to zero to simulate the lack of observation during gaps. Ideally any gap should be sufficient far away from reference time $t_c$ in order to avoid loss of detection entirely~\cite{Dey_2021}, however in a separate case we would also place a long gap arbitrarily close to $t_c$ in order to test extreme cases for our treatment strategy.

\subsubsection{Windowing}
In order to counter the spectral leakage introduced by gaps, it is a simple but effective approach to apply a windowing function to the gap, and both sides of the gaps smoothed by a windowing function are defined as below~\cite{Dey_2021},
\begin{equation}
w(t)= \begin{cases}\frac{1}{2}\left(1+\cos \left[\pi\left(\frac{t-t_{\mathrm{s}}-t_{\mathrm{tr}}}{t_{\mathrm{tr}}}\right)\right]\right) & \text { for } t_{\mathrm{s}}-t_{\mathrm{tr}}<t<t_{\mathrm{s}}, \\ 0 & \text { for } t_{\mathrm{s}}<t<t_{\mathrm{e}}, \\ \frac{1}{2}\left(1+\cos \left[\pi\left(\frac{t-t_{\mathrm{e}}-t_{\mathrm{tr}}}{t_{\mathrm{tr}}}\right)\right]\right) & \text { for } t_{\mathrm{e}}<t<t_{\mathrm{e}}+t_{\mathrm{tr}},\\ 1 & \text {otherwise}.\end{cases}
\end{equation}
As is shown in previous works~\cite{Dey_2021}, the empirical transition time for 3-day-long gap was 2 days on each side, we retained this numerical setting for $t_{tr}$ as default and this windowing technique will serve as a baseline for gap treatment later displayed as Gapped.

\subsubsection{Kalman Smoother}
Kalman filter was developed independently by Swerling~\cite{swerling1958proposed} and Kalman~\cite{Kalman1960}, and has seen extensive use since its original use in the Apollo Navigation system~\cite{McGee:1985ne}. Kalman filter performs well in case where observation points are missing and predictions on these missing points as well as system states are needed. In order to accomplish this, we need both the forward and backward recursions, known as Kalman filter and Kalman smoother respectively. The definitions for the Kalman filter and the Kalman smoother go as follows.

We consider a linear time-invariant dynamical system (LDS),
\begin{equation}
\begin{aligned}
\mathbf{x}_{t+1} &= A \mathbf{x}_t + \mathbf{w}_t, \
\mathbf{y}_t &= C \mathbf{x}_t + \mathbf{v}_t,
\end{aligned}
\end{equation}
where $\mathbf{w}_t \sim \mathcal{N}(0, Q)$ and $\mathbf{v}_t \sim \mathcal{N}(0, R)$. The initial state is distributed as $\mathbf{x}_1 \sim \mathcal{N}(\boldsymbol{\pi}_1, V_1)$.
Let $\mathbf{x}_t^\tau = \mathbb{E}[\mathbf{x}_t \mid {\mathbf{y}}_1^\tau]$ and $V_t^\tau = \operatorname{Var}[\mathbf{x}_t \mid {\mathbf{y}}_1^\tau]$ denote the filtered ($\tau = t$) or smoothed ($\tau = T$) posterior estimates.
The Kalman Filter is defined as follows.
For time update (prediction) we have the following:
\begin{equation}
\begin{aligned}
\mathbf{x}t^{t-1} &= A \mathbf{x}{t-1}^{t-1}, \\
V_t^{t-1} &= A V_{t-1}^{t-1} A^\top + Q.
\end{aligned}
\end{equation}
And for measurement update (correction) we have:
\begin{equation}
\begin{aligned}
K_t &= V_t^{t-1} C^\top (C V_t^{t-1} C^\top + R)^{-1}, \\
\mathbf{x}_t^t &= \mathbf{x}_t^{t-1} + K_t (\mathbf{y}_t - C \mathbf{x}_t^{t-1}), \\
V_t^t &= (I - K_t C) V_t^{t-1}.
\end{aligned}
\end{equation}
The recursion is initialized with $\mathbf{x}_1^0 = \boldsymbol{\pi}_1$ and $V_1^0 = V_1$.

For the backwards step, the Rauch–Tung–Striebel (RTS) smoother computes smoothed estimates as:
\begin{equation}
\begin{aligned}
J_{t} &= V_t^t A^\top (V_{t+1}^{t})^{-1}, \\
\mathbf{x}t^T &= \mathbf{x}t^t + J_t (\mathbf{x}{t+1}^T - \mathbf{x}{t+1}^{t}), \\
V_t^T &= V_t^t + J_t (V_{t+1}^T - V_{t+1}^{t}) J_t^\top.
\end{aligned}
\end{equation}
This recursion proceeds backward from $t = T-1$ to $1$, starting with $\mathbf{x}_T^T$ and $V_T^T$ from the final filter step. For further details, see the following technical report~\cite{Yu_derive} for a more comprehensive derivation.

In practice, we made use of the existing library of pykalman~\cite{pykalman} to implement both Kalman filter and smoother. For our case, we made the following approximation that in a gap that is not too long and not too close to $t_c$, it is assumed that a Linear-Gaussian model can describe the data accurately enough for our analysis. Thus, we gave the following initial parameters for our Kalman filter as is shown in Table~\ref{tab:initialParams}.

\begin{table}
\centering
\begin{tabular}{ll} 
\hline
Parameter Name & Notation \\
\hline
initial state mean & $\mu_0$ \\
initial state covariance & $\Sigma_0$ \\
transition matrices & $A$ \\
transition offsets & $b$ \\
transition covariance & $Q$ \\
observation matrices & $C$ \\
observation offsets & $d$ \\
observation covariance & $R$ \\
\hline
\end{tabular}
\caption{Notations for model parameters used in the state-space formulation. Note it is easy to give an a priori setting for these from our previous knowledge of the noise characteristics and observation.}
\label{tab:initialParams}
\end{table}

During our construction of the transition matrix, we use the Taylor expansion to capture the nonlinearity of the model locally, the definition for the three-dimensional state vector goes as follows,
\begin{equation}
h(t + \Delta t) = h(t) + \dot{h}(t)\, \Delta t + \tfrac{1}{2} \ddot{h}(t)\, \Delta t ^2.
\end{equation}
Then the corresponding transition matrix $A$ can be structured as follows,
\begin{equation}
A =
\begin{bmatrix}
1 & \Delta t & \tfrac{1}{2} \Delta t ^2\\
0 & 1 & \Delta t \\
0 & 0 & 1
\end{bmatrix}.
\end{equation}
We also utilize the built-in Expectation-Maximization (EM) algorithm to iteratively optimize the likelihood of the observed measurements. In our case, it is chosen to optimize over both transition covariance $Q$ and observation covariance $R$ in order to dynamically adjust Kalman filter's behavior. We empirically set number of iterations for EM algorithm to five in order to ensure convergence and to minimize the computational cost required. It is noted that further iterations can, in some cases, lead to a minor improvement in posterior estimation performance, yet 5 is a safe default setting. Data points from smoothed observations then replace respective empty data points in gaps to form a repaired version of the gapped observation that fits for further Bayesian Analysis.

\subsubsection{Bayesian Analysis}
On the basis of our analysis lies the Bayesian Theorem,
\begin{equation}
p(\theta|d)=\frac{p(d|\theta) p(\theta)}{p(d)}
\end{equation}
where $p(\theta|d)$ is the posterior probability distribution, $p(d|\theta)$ is the likelihood, $p(\theta)$ is the prior probability distribution and $p(d)$ is the evidence. Here the evidence $p(d)$ stays constant for the same set of data throughout our work and is treated as stationary.

For a stationary and Gaussian noise, we have the likelihood function written as follows,
\begin{equation}
\begin{aligned}
&\mathcal{L}(\theta)=\mathcal{A} \cdot \exp \left[-\frac{1}{2}(\mathbf{d}-\mathbf{h}(\theta) \mid \mathbf{d}-\mathbf{h}(\theta))\right],
\end{aligned}
\end{equation}
where $\mathcal{A}$ is the normalization constant and the inner product is defined as
\begin{equation}
(a \mid b)=4 \operatorname{Re} \int_0^{\infty} \frac{\tilde{a}^*(f) \tilde{b}(f)}{S_n(f)} \mathrm{d} f.
\end{equation}
Here $S_n$(f) is given by the noise PSD.

As was pointed out in Ref.~\cite{Dey_2021}, calculation of the loglikelihood function of a point in the parameter space is computational expensive which involves transforming signal into time domain in order to place gaps and apply treatment procedures before converting back to frequency domain for analysis. This requires a robust sampling technique to achieve convergence and in our case we chose the emcee based sampler Eryn~\cite{2013PASP..125..306F, Karnesis_2023, michael_katz_2023_7705496} in order to achieve the Ensembled Markov Chain Monte Carlo (MCMC) sampling required for out posterior estimation. Eryn also includes partial support for GPU-acceleration that can be incorporated with the GPU-accelerated template generation from bbhx which in turn speed-up the parameter estimation process.

\section{Results}
\subsection{Injection Parameters}
Here we chose to inject the same set of parameters from the Heavy case in this previous work~\cite{Dey_2021}, this case was also the set of parameters used by another separate work in 2021~\cite{Katz_2022} and has its origin in the LISA Data Challenge data set, so it is a good basis for an analysis for gap treatment. The inject parameters are shown in Table~\ref{tab:injection_params}.

\begin{table}
\centering
\begin{tabular}{lll}
\hline
\textbf{Parameter} & \textbf{Symbol} & \textbf{Value} \\
\hline
Chirp mass & $\mathcal{M}_c$ [$M_\odot$] & 1,543,972.48 \\
Mass ratio & $q$ & 0.478 \\
Primary spin & $a_1$ & 0.75 \\
Secondary spin & $a_2$ & 0.62 \\
Coalescence time & $t_c$ [seconds] & 24960000 \\
Coalescence phase & $\phi_c$ [rad] & 1 \\
Luminosity distance & $D_L$ [Mpc] & 56,005.78 \\
Inclination angle & $\iota$ [rad] & 1.22 \\
Ecliptic longitude & $\lambda$ [rad] & 3.51 \\
Ecliptic latitude & $\beta$ [rad] & 0.29 \\
Polarization angle & $\psi$ [rad] & 2.94 \\
\hline
\end{tabular}
\caption{Injected binary black hole parameters used for the gravitational wave simulation. Note that due to difference in precision, the actual injection used in previous works~\cite{Dey_2021,Katz_2022} varied slightly.}
\label{tab:injection_params}
\end{table}

\begin{figure}
\centering
\includegraphics[width=0.5\textwidth]{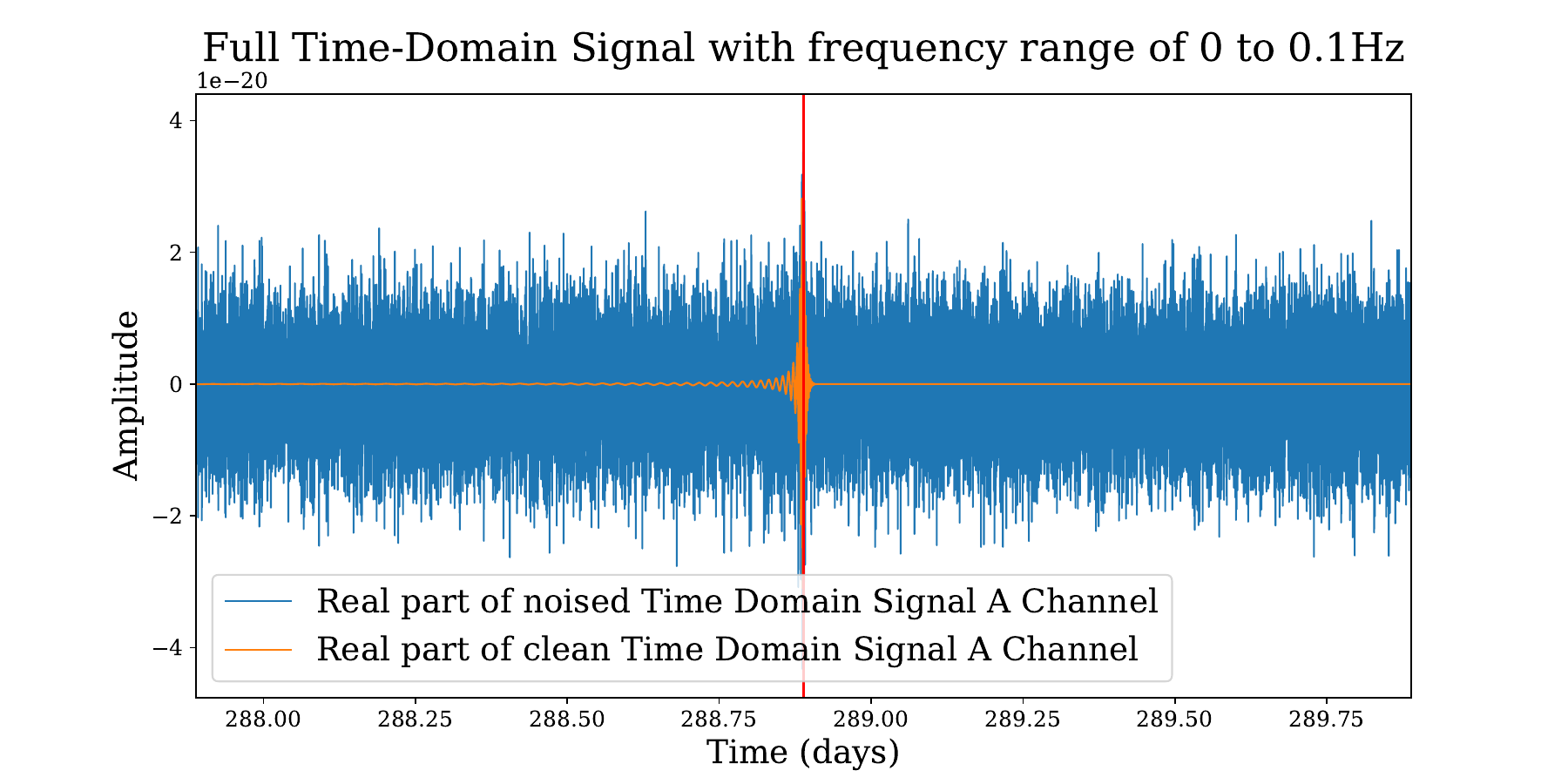}
\caption{Injected GW signal with parameters in Table \ref{tab:injection_params} into noise generated from PSD.}
\label{fig:mygraph}
\end{figure}

\begin{figure}
\centering
\includegraphics[width=0.5\textwidth]{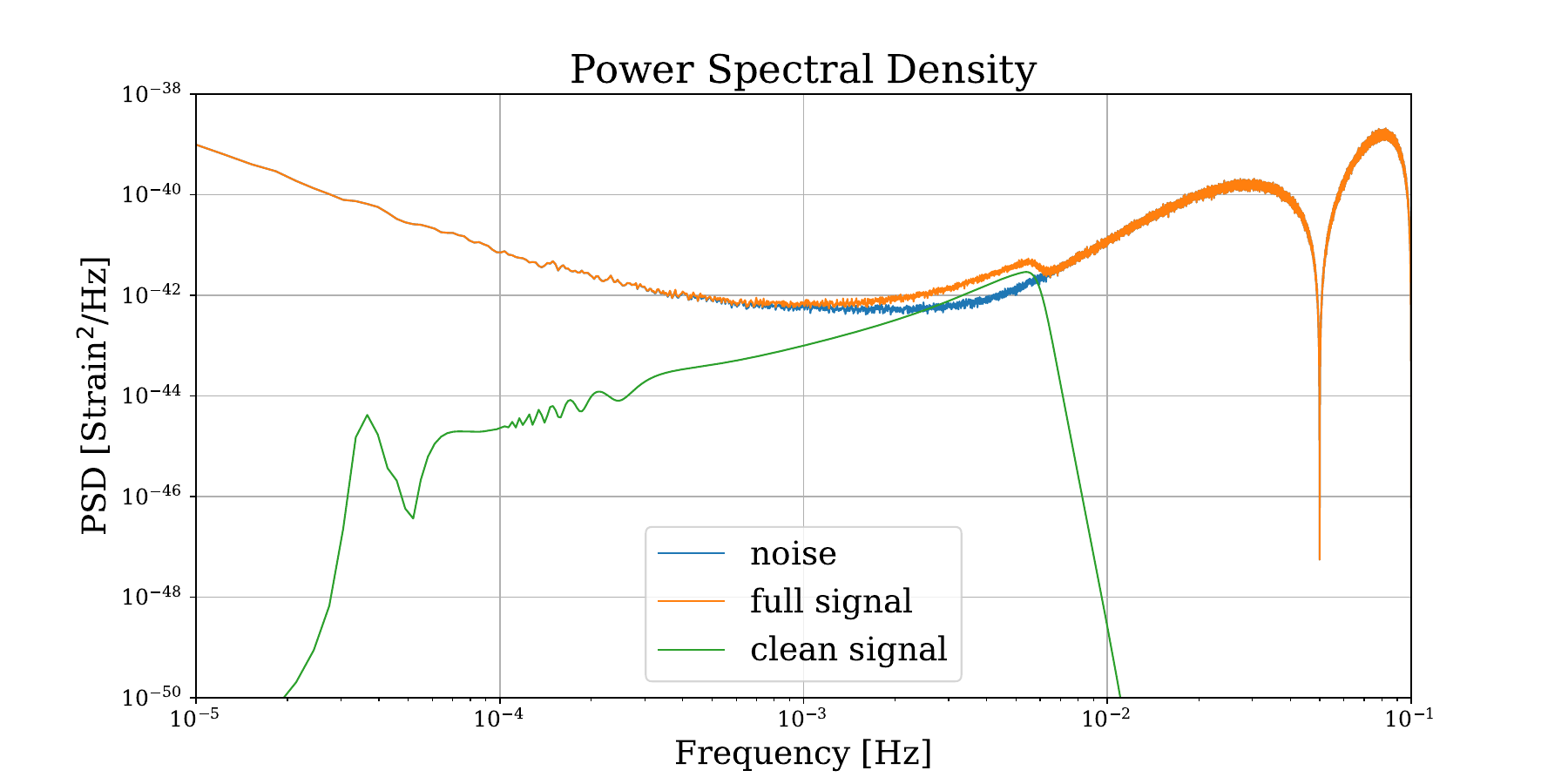}
\caption{PSD estimation for noise, pure signal and noised signal respectively, one can clearly see the signal reached out of the noise background.}
\label{fig:PSD}
\end{figure}

One can see from Fig.~\ref{fig:mygraph} that the injected signal in time domain over a noise background with the red line indicates the reference time $t_c$, this will serve as a basis for later gap insertion and treatment as well as respective parameter estimation processes for optimal (which means full data without gaps), gapped (with gaps but smoothed with windowing function) and repaired (Kalman-smoother repaired data). The PSDs for this case can be seen in Fig.~\ref{fig:PSD}.

At its present shape the original signal is prepared and is referred to as Optimal Data later in the posterior estimation results.

\subsection{Results}
The gags are introduced through the random number generator which determines both gap positions and gap lengths, precautions were made here to prevent the merging of overlapping gaps into larger gaps. This ensured the actual gaps introduced and treated as intended (especially in case of traditional windowing methods).
In our case, we inserted gaps following the pattern in Table~\ref{tab:gap_pattern}.

\begin{table}
\centering
\begin{tabular}{lll}
\hline
\textbf{Parameter} & \textbf{Value} \\
\hline
Gap duration mean & 600 s  \\
Gap duration std. dev. & 60 s \\
Gap interval mean & 8,640 s (0.1 days) \\
Gap interval std. dev. & 1,728 s (0.02 days) \\
\hline
\end{tabular}
\caption{Parameters defining the synthetic gap pattern.}
\label{tab:gap_pattern}
\end{table}

\begin{figure*}[t]
    \centering
    \subfigure[Comparision between posterior estimation results for intrinsic parameters from repaired and gapped data]{\includegraphics[width=0.49\linewidth]{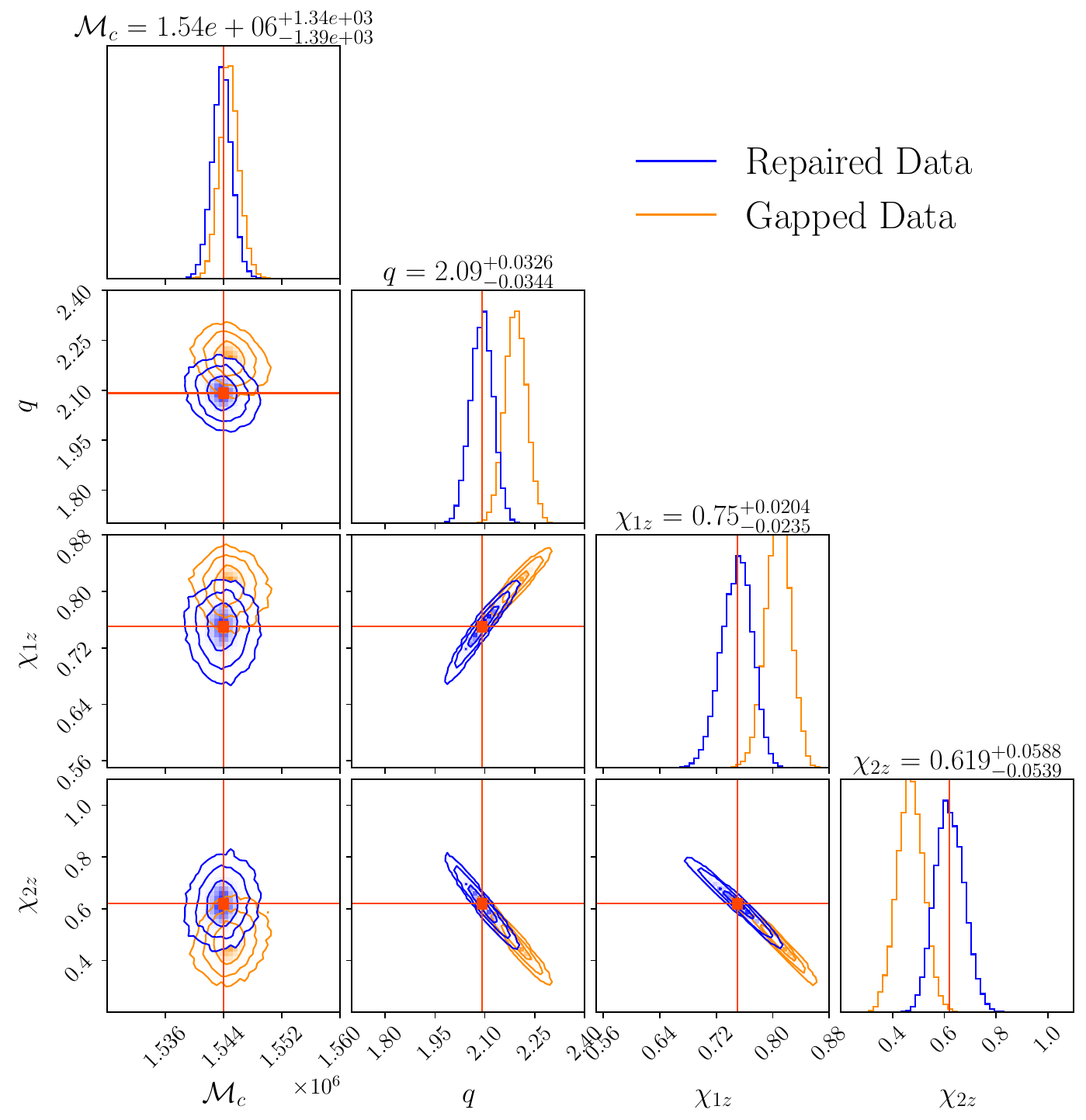}}
    \hfill
    \subfigure[Comparision between posterior estimation results for intrisic parameters from repaired and optimal data]{\includegraphics[width=0.49\linewidth]{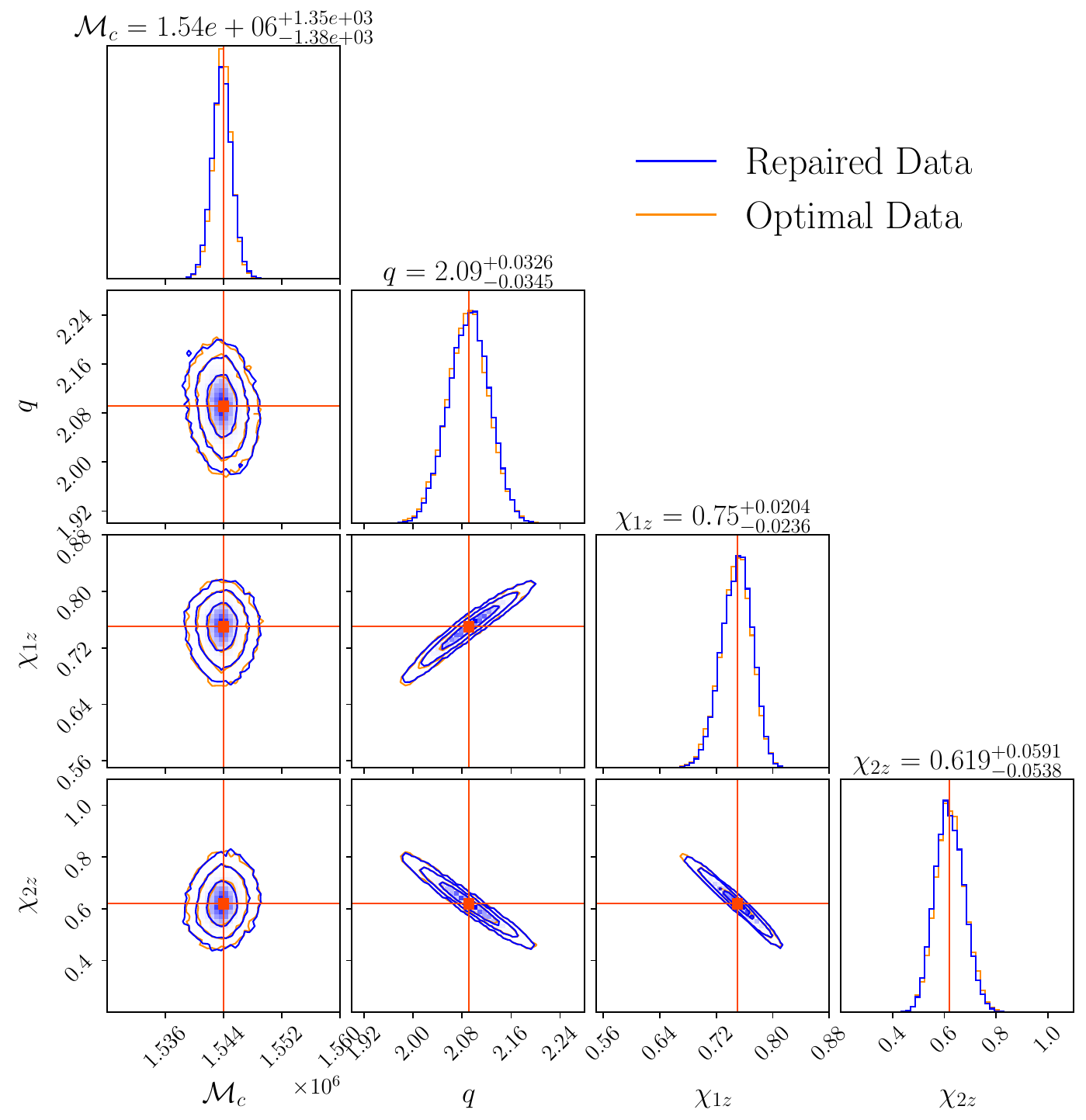}}
    \caption{As can be seen from the results on the left, the above mentioned gap in Table \ref{tab:gap_pattern} introduced a significant bias in the intrinsics parameters especially in mass ratio $q$ and spins $\chi_{1z}$ and $\chi_{2z}$ with the injected value is almost $3\sigma$   away from the posterior mean, the repaired value shows mitigation for such bias and shows a similar posterior as the optimal data (the original data with no gaps).}
    \label{fig:vs}
\end{figure*}

Considering LISA Pathfinder mission observed one event per day that led to data loss (in this case mainly glitches)~\cite{PhysRevLett.120.061101}, and we increased this rate tenfold in order to observe a clearer effect on posterior estimation results. Each gap lasts 10 minutes on average and a total loss cap of 1\% is introduced to avoid too much data loss in this aggressive gap setting. The parameter estimation outputs are shown in Fig.~\ref{fig:vs}, demonstrating that our Kalman smoother reparation corrected the bias and broadening of posterior distribution introduced by data gaps. This highlights the superiority Kalman smoother has over simple windowing technique, as it not only repairs the missing noise but also attempts to recover some aspects of the signal by making predictions based on nearby surviving observations rather than discarding even more data points to address the spectral leakage problem.

\begin{figure*}[t]
    \centering
    \includegraphics[width=1\linewidth]{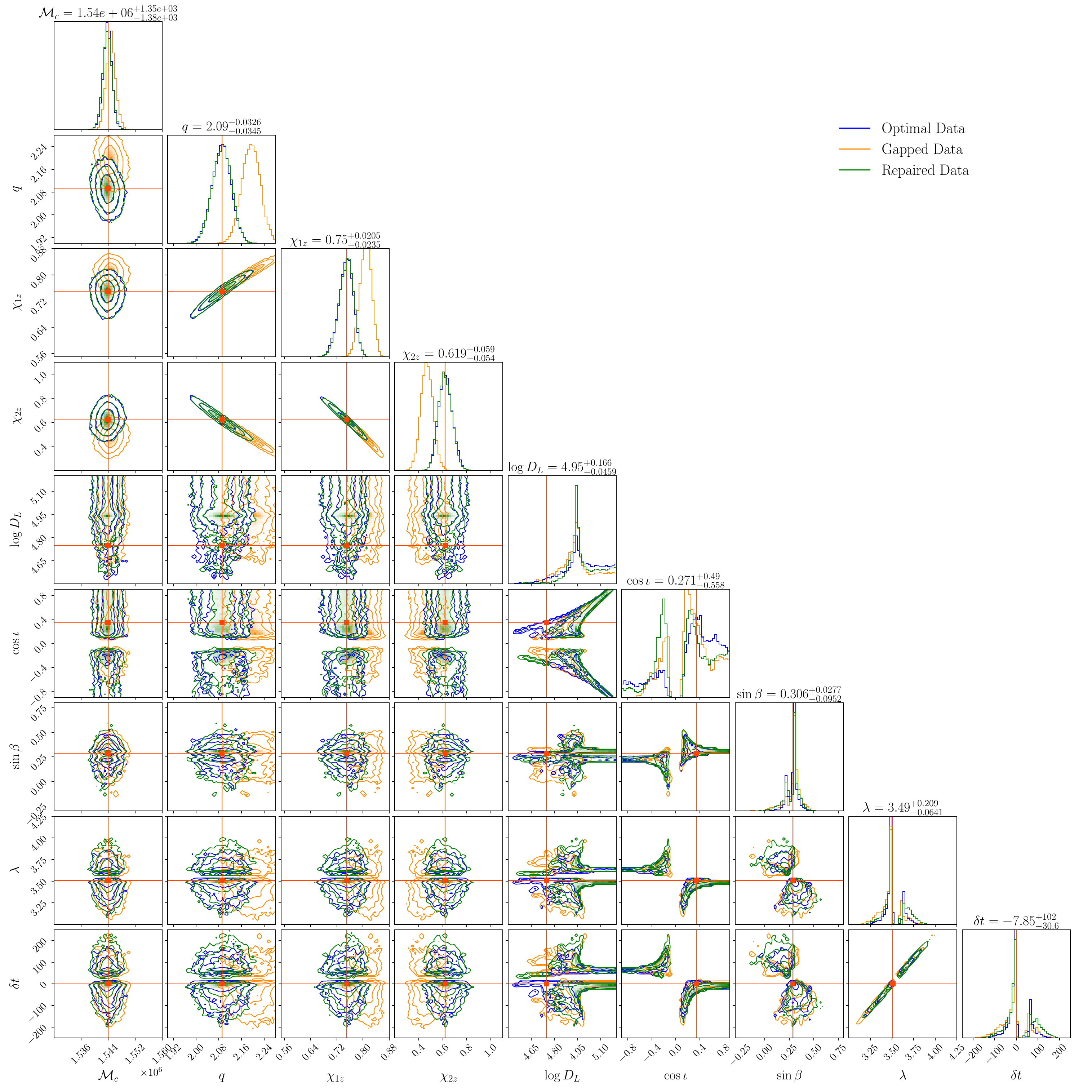}
    \caption{Results of 9-dimension parameter estimation for the Optimal, Gapped and repaired data. Note here $\delta t$ stands for the deviation of estimated $t_c$ from injected $t_c$ value. One can see that both the Optimal and the Repaired case retain the same posterior shape for extrinsic parameters like luminosity distance $D_L$, inclination angle $\iota$, ecliptic longitude $\lambda$ and ecliptic latitude $\beta$.}
    \label{fig:full}
\end{figure*}

Comparisons to the optimal data case as is shown in Fig.~\ref{fig:vs}, shows that even in this particular gap setting, the repaired data still has an unbiased posterior estimation close to the optimal case. The 9-dimension parameter estimation result is given in Fig.~\ref{fig:full}, with optimal data in blue, gapped data in orange and repaired data in green, which shows that for extrinsic parameters the posterior has similar range and multi-modality as is also observed in previous works with the same set of data~\cite{Dey_2021,Katz_2022}.

Our data generation, imputation and parameter estimation were conducted on a dual Intel Xeon Platinum 8268 CPU server (48 cores, 96 threads) with 2.9 GHz base clock speed, and a NVIDIA A800 graphics card with 80 GB VRAM. The imputation itself took several thousands of seconds, which is a fraction of time needed compared to the total time needed for a full parameter estimation. A reduced version (which optimized for less VRAM consumption) was also run on a personal laptop with a NVIDIA RTX4070 Laptop Graphics card with no significant performance degradation for data generation and imputation process, further proving the robustness of our procedure.

\section{Conclusions}
In this work we propose a new strategy based on Kalman filtering and smoothing in order to mitigate the effects that data gaps have on the parameter estimation of MBHB signals. Our strategy fills in missing data by imputing predictions derived from surrounding observations using Kalman filtering and smoothing. It is  demonstrated that when handling our test scenario where gaps lasted around 15 minutes long and took up up to 1\% of total data points, the Kalman Smoother performs significantly better than windowing techniques, as it utilizes more data points around the gaps instead of discarding them to reduce spectral leakage.

Another highlight of our method's performance is that it requires minimal prior knowledge to start our imputation process, we only utilize the information that the signal is chirping and the noise level is estimated dynamically for both transition covariance $Q$ and observation covariance $R$. This requires less computation cost compared to the iterative estimation of both signal and noise parameters used in previous noise inpainting methods~\cite{Blelly_2021,Wang_2025} as is shown in Table~\ref{tab:complexity_comparison}. 

\begin{table}
\centering
\begin{tabular}{lcc}
\hline\hline
\textbf{Method} & \textbf{Computational Complexity} \\
\hline
Kalman Smoother & $\mathcal{O}(N n^3 )$ \\
Noise Inpainting & $\mathcal{O}(N \log N )+\mathcal{O}(M^2)$ \\
\hline\hline
\end{tabular}
\caption{Comparison of computational complexity between Kalman smoothing and noise inpainting techniques for each iteration. Here, $N$ is the number of data points, $M$ is the number of missing points, $n$ is the state vector dimension and in our case $n = 3$. For noise inpainting, $\mathcal{O}(N \log N )$ stands for the loglikelihood and $\mathcal{O}(M^2)$ for the inpainting~\cite{Wang_2025}.}
\label{tab:complexity_comparison}
\end{table}

Note that for scenarios where both the number of data points and lost data points are small, both techniques have a theoretically comparable computational complexity. However, the complexity of noise inpainting increases significantly with increasing data size, as it scales quadratically with the number of missing data points. In contrast, the Kalman smoother scales linearly with the total number of time steps and cubically with the (typically low-dimensional) state vector size. As a result, for longer signals or higher-resolution datasets, the Kalman-based method becomes substantially more computationally efficient.

\begin{figure}[!htbp]
\centering
\includegraphics[width=0.5\textwidth]{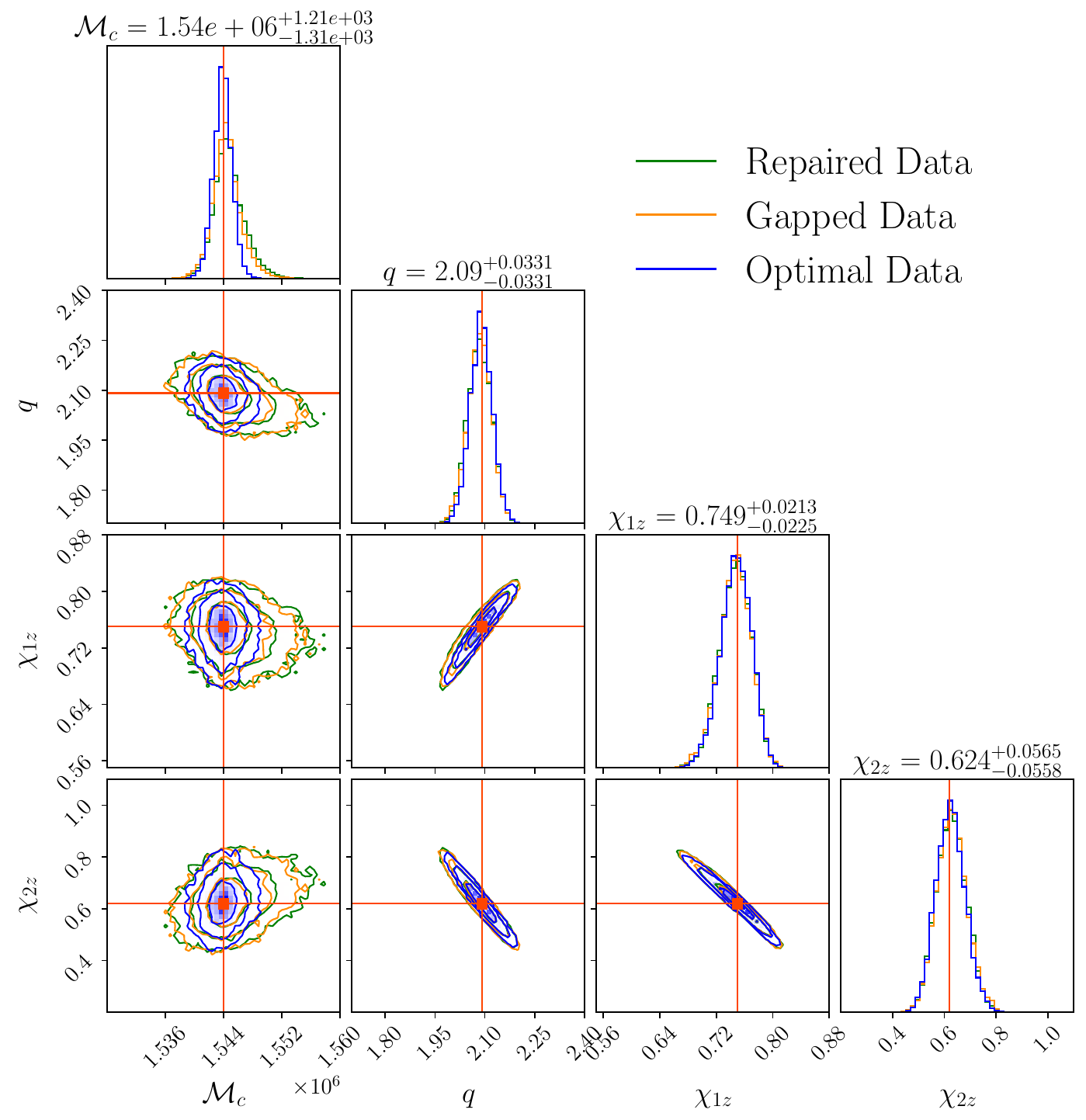}
\caption{Comparison of parameter estimation results for single 3-day long gap located 0.2 days before $t_c$. Note that in this case only one gap is introduced and the total data loss ratio is 0.8\% (3 days out of one year of observation).}
\label{fig:long}
\end{figure}

However, for long gaps that are very close to reference time $t_c$, the Kalman-smoother based treatment shows significant performance degradation where the repaired data yielded little improvemet in parameter estimation results and lead to a broader posterior for intrinsic parameters, as can be seen in Fig~\ref{fig:long}. One can see that with only one 3-day long gap located near reference time $t_c$, the posterior distribution of chirp mass $\mathcal{M}_c$ displays a slight bias together with a noticeable widening that the Kalman Filter and Smoother cannot handle. This might arise from not having enough valid data points on both ends of the gap to make a good prediction of lost data points. Also, gaps that are too long tend to be hard to fill as the predictions made by Kalman Filter alone does not venture deep into gaps and only peripheral points near the edge of the gap is imputed. This can certainly be a future focus for improvement for a more comprehensive gap treatment scheme.

On the other hand, the Taylor expansion approximation as well as the linearity assumption used when implementing Kalman Filter can be faulty for long gaps and portions of the observation where the signal frequency changes rapidly. The Kalman filter and smoother rely solely on the states of the observed points to predict the states of missing ones. If the signal exhibits strong nonlinearity or rapid frequency changes, the linear model may fail to capture its dynamics accurately. As a result, the predicted states may provide limited information for posterior estimation, and in extreme cases, may even introduce additional bias or uncertainty.

These limitations call for a native nonlinear Kalman Filtering algorithm to address such issues. Nonlinear Kalman Filter variants such as the Unscented Kalman Filter (UKF) and the Extended Kalman Filter (EKF) propagate mean and covariance through nonlinear transformations, enabling more accurate tracking in the presence of nonlinear signal behaviors. However, configuring and tuning nonlinear Kalman filters can be nontrivial, often requiring careful modeling of the system dynamics and noise characteristics. We leave the exploration of these more advanced nonlinear filters to future work.

\begin{acknowledgments}
We thank Minghui Du for insightful discussions and assistance with time-delay interferometry and noise generation. We also want to acknowledge Kallol Dey for his valuable advice on parameter estimation. 
This research is funded by the Strategic Priority Research Program of the Chinese Academy of Sciences under Grant No. XDA15021100, as well as the Fundamental Research Funds for the Central Universities.
\end{acknowledgments}

\nocite{*}
\bibliographystyle{apsrev4-2}
\bibliography{ref}

\end{document}